\documentclass[sigconf]{acmart}
\usepackage{blindtext}
\usepackage{booktabs}
\usepackage{graphicx}
\usepackage{blindtext}
\usepackage{tabularx}

\renewcommand\footnotetextcopyrightpermission[1]{}
\hypersetup{breaklinks=true}
\begin{document}

\title{Understanding Car-Speak:\\Replacing Humans in Dealerships}

\author{Habeeb Hooshmand and James Caverlee}
\email{(habeebh, caverlee)@tamu.edu}
\affiliation{Department of Computer Science and Engineering \\ 
Texas A\&M University, USA}

\renewcommand{\shortauthors}{Hooshmand and Caverlee}

\begin{abstract}

A large portion of the car-buying experience in the United States involves interactions at a car dealership. At the dealership, the car-buyer relays their needs to a sales representative. However, most car-buyers are only have an abstract description of the vehicle they need. Therefore, they are only able to describe their ideal car in ``car-speak''. Car-speak is abstract language that pertains to a car's physical attributes. In this paper, we define car-speak. We also aim to curate a reasonable data set of car-speak language. Finally, we train several classifiers in order to classify car-speak. 

\end{abstract}

\maketitle

\section{Introduction}\label{sec:intro}
A large portion of the car-buying experience in the United States involves interactions at a car dealership~\cite{deloitte2014, kershner2010, barley2015}. Traditionally, a car dealer listens and understands the needs of the client and helps them find what car is right based on their needs. 

With the advent of the internet, many potential car buyers take to the web to research cars before going to a dealership in person~\cite{deloitte2014, barley2015}. However, nearly 50\% of customers bought a car at the dealership based on the sales representative's advice, not their own research~\cite{kershner2010, barley2015}. 

Throughout this interaction the dealer is acting as a type of translator or classifier. The dealer takes a natural language input (e.g. ``I need a fast, family friendly, reliable car under \$20k'') and returns a list of suggestions. The dealer understands the ideas of ``fast'', ``family friendly'', and ``reliable'' and is able to come up with a reasonable recommendation based on this knowledge.

In this paper we aim to create a system that can understand car-speak based on some natural language input (we want to recreate the dealer from above). But how do we prepare a proper training set for a Natural Language model? What model is best suited to this problem? Can this model take a human out of the car-buying process? To answer these questions, the remainder of this paper makes the following contributions:
\begin{itemize}
    \item Defining ``car-speak'' and its role in the car-buying process.
    \item Appropriate training data for a Natural Language model.
    \item A model that is able to properly classify car-speak and return a car. 
\end{itemize}
We aim to accomplish these goals in a scientific manner, using real data and modern methods. 

\begin{figure}[t]
    \centering
    \includegraphics[width=0.30\textwidth]{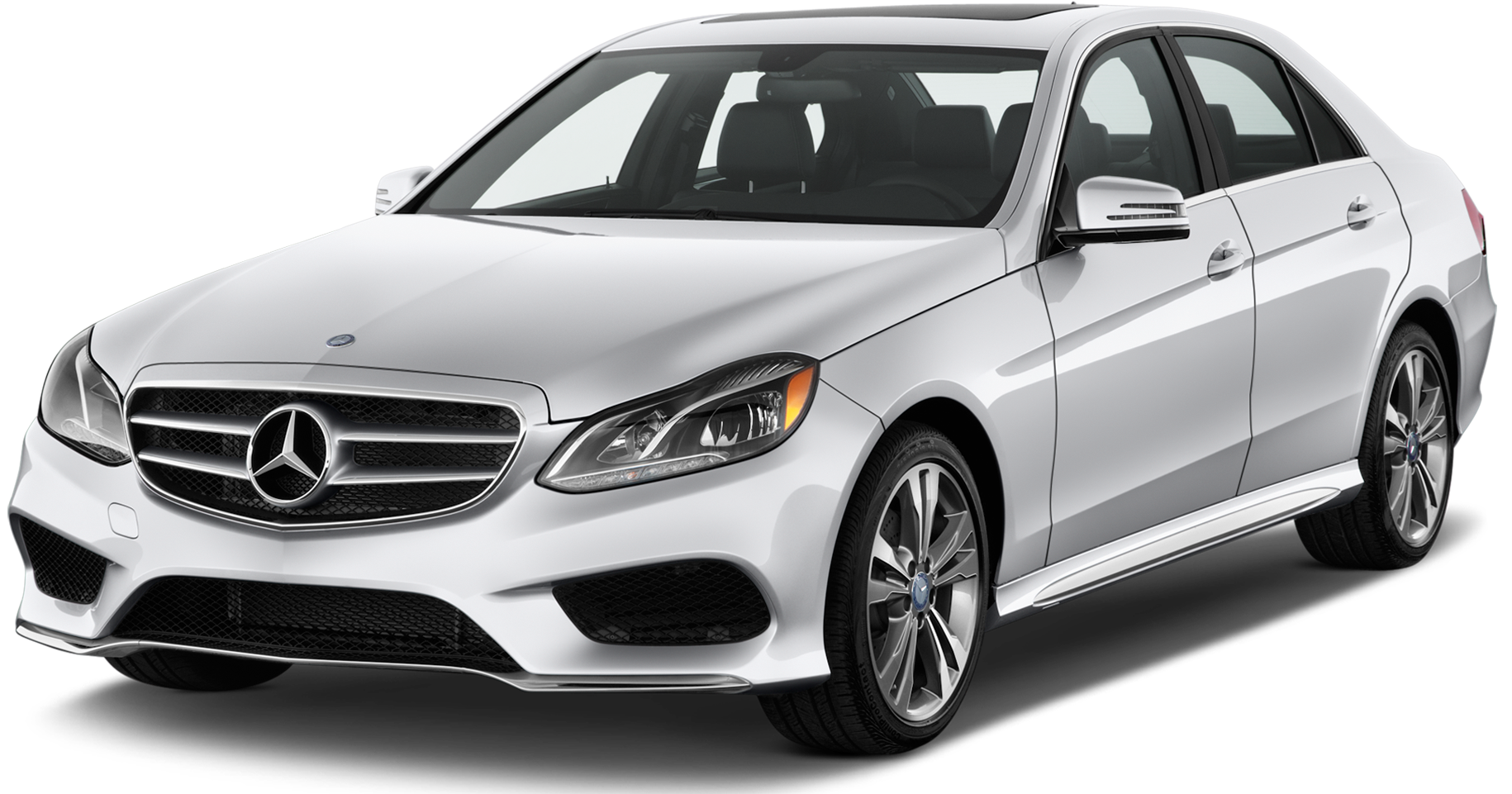}
    \includegraphics[width=0.35\textwidth]{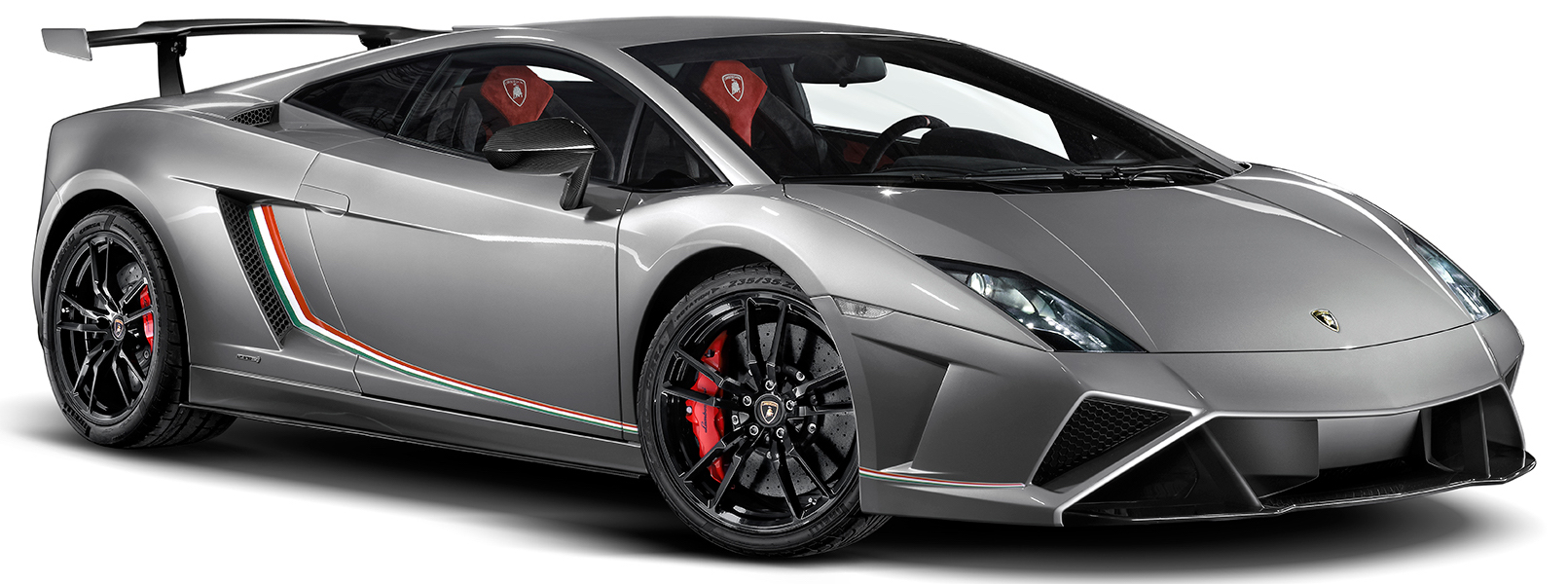}
    \caption{Both of these cars can achieve high speeds. Which is ``fast''?}
    \label{fig:car-comparison}
\end{figure}

\newpage
\section{Related Work}\label{sec:related_work}
There has been some work done in the field of car-sales and dealer interactions. However, this is the first work that specifically focuses on the 

Deloitte has published a report on the entire car-buying process~\cite{deloitte2014}. The report goes into great depth about the methods potential buyers use to find new cars to buy, and how they go about buying them. The report tells us that there are several unique phases that a potential buyer goes through before buying a car. 

Verhoef et al. looked at the specifics of dealer interaction and how dealers retain customers~\cite{verhoef2007}. Verhoef tells us how important dealers are to the car-buying process. He also explains how influential a dealer can be on what car the buyer purchases.

Jeff Kershner compiled a series of statistics about Dealership Sales~\cite{kershner2010}. These statistics focus on small social interactions~\cite{hare1965} between the dealer and the buyer.

Barley explains the increasing role of technology in the car-buying process~\cite{barley2015}. Barley tells us that users prefer to use technology/robots to find the cars they want to buy instead of going to a dealer, due the distrust towards sales representatives.

\section{What is Car-speak?} \label{sec:car-speak}

When a potential buyer begins to identify their next car-purchase they begin with identifying their needs. These needs often come in the form of an abstract situation, for instance, ``I need a car that goes really fast''. This could mean that they need a car with a V8 engine type or a car that has 500 horsepower, but the buyer does not know that, all they know is that they need a ``fast'' car. 

The term ``fast'' is car-speak. Car-speak is abstract language that pertains to a car's physical attribute(s). In this instance the physical attributes that the term ``fast'' pertains to could be the horsepower, or it could be the car's form factor (how the car looks). However, we do not know exactly which attributes the term ``fast'' refers to. 


The use of car-speak is present throughout the car-buying process. It begins in the \textit{Research Phase} where buyers identify their needs~\cite{deloitte2014}. When the buyer goes to a dealership to buy a car, they communicate with the dealer in similar car-speak~\cite{barley2015} and convey their needs to the sales representative. Finally, the sales representative uses their internal classifier to translate this car-speak into actual physical attributes (e.g. `fast' $ \longrightarrow $ `700 horsepower \& a sleek form factor')  and offers a car to the buyer.

Understanding car-speak is not a trivial task. Figure~\ref{fig:car-comparison} shows two cars that have high top speeds, however both cars may not be considered ``fast''. We need to mine the \textit{ideas} that people have about cars in order to determine which cars are ``fast'' and which cars are not. 


\section{Gathering Car-speak Data}

We aim to curate a data set of car-speak in order to train a model properly. However, there are a few challenges that present themselves: What is a good source of car-speak? How can we acquire the data? How can we be sure the data set is relevant? \\

\begin{table}[t]
    \caption{Excerpts from car reviews.}
    \label{table:quotes}
    \begin{tabular}{| c | c |}
        \hline  
        \textbf{Car} & \textbf{Excerpt}  \\ 
        \specialrule{.3em}{.2em}{.2em}
        Acura ILX & ``making it one of the \\ 
        & most \textbf{efficient} cars'' \\ 
        \hline
        Audi A6 & ``best cars for \textbf{families}'' \\ 
        \hline
        Chevrolet Impala & ``strong mix of \textbf{comfort} and \\ 
        & \textbf{safety} features'' \\
        \hline
        Lexus ES & ``\textbf{luxurious} cabin \textbf{comfortable} seats'' \\
        \hline
        Mercedes-Benz S & ``\textbf{handles like a sports sedan} \\ 
        & despite its large size'' \\
        \hline
        Toyota Camry & ``top-notch \textbf{reliability} and \\
        & a \textbf{good value} proposition'' \\
        \hline
        
    \end{tabular}
\end{table}

\noindent\textbf{What is a good source of car-speak?} We find  plenty of car-speak in car reviews. Table~\ref{table:quotes} provides excerpts from reviews with the car-speak terms \textbf{bolded}. Car reviews often describe cars in an abstract manner, which makes the review more useful for car-buyers. The reviews are often also about specific use-cases for each car (e.g. using the car to tow a trailer), so they capture all possible aspects of a car. The reviews are each written about a specific car, so we are able to map car-speak to a specific car model. 

We choose the reviews from the U.S. News \& World Report because they have easily accessible full-length reviews about every car that has been sold in the United States since 2006~\cite{usnews}. \\

\noindent\textbf{How can we acquire the data?} We can acquire this data using modern web-scraping tools such as beautiful-soup. The data is publicly available on \url{https://cars.usnews.com/cars-trucks}~\cite{usnews}. These reviews also include a scorecard and justification of their reviews. \\

\noindent\textbf{How can we be sure the data set is relevant?} On average vehicles on United States roads are 11.6 years old, making the average manufacturing year 2006-2007~\cite{bts2018, hirsch2014}. In order to have a relevant data set we gather all of the available reviews for car models made between the years 2000 and 2018.



\begin{figure}[t]
    \centering
    \includegraphics[width=0.45\textwidth]{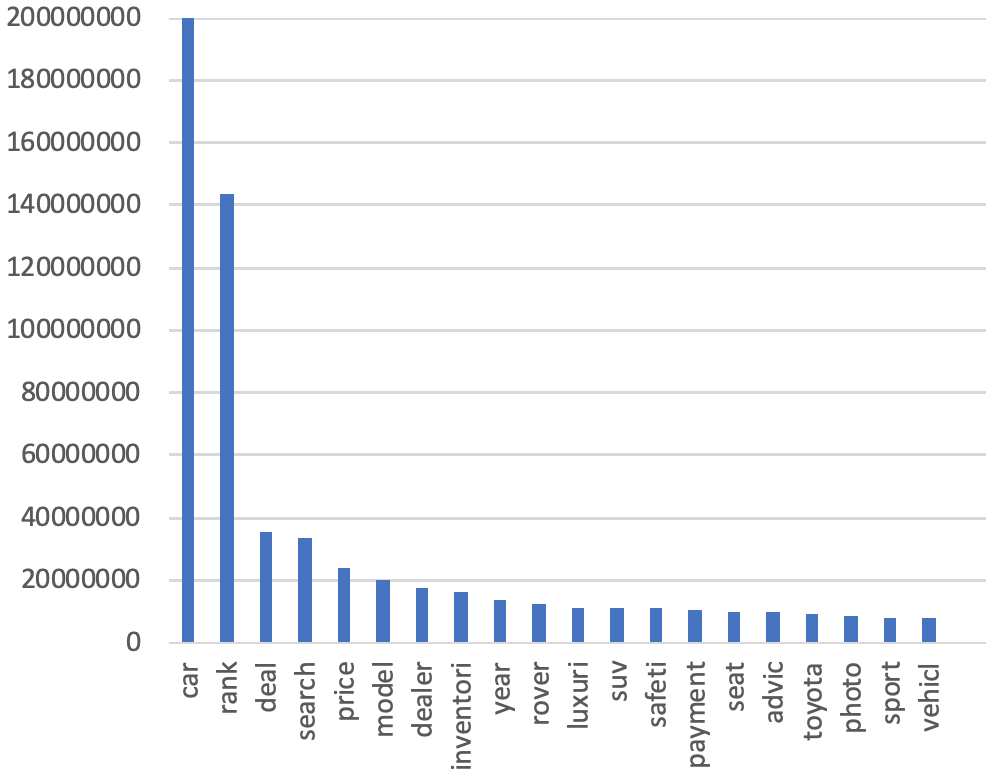}
    \caption{The frequencies of the top 20 words in reviews.}
    \label{fig:filtered_distro}
\end{figure}

\section{Translating Car-Speak}

Our data set contains $3,209$ reviews about $553$ different cars from $49$ different car manufacturers. In order to accomplish our goal of translating and classifying car-speak we need to filter our data set so that we only have the most relevant terms. We then need to be able to weight each word in each review, so that we can determine the most relevant ideas in each document for the purpose of classification. Finally, we need to train various classification models and evaluate them. 

\subsection{Filtering the Data}

We would like to be able to represent each car with the most relevant car-speak terms. We can do this by filtering each review using the NLTK library~\cite{bird2009}, only retaining the most relevant words. First we token-ize each review and then keep only the nouns and adjectives from each review since they are the most salient parts of speech~\cite{hurlburt1954}. This leaves us with $10,867$ words across all reviews. Figure~\ref{fig:filtered_distro} shows the frequency of the top 20 words that remain.

Words such as ``saftey'' and ``luxury'' are among the top words used in reviews. These words are very good examples of car-speak. Both words are abstract descriptions of cars, but both have physical characteristics that are associated with them as we discussed in Section~\ref{sec:car-speak}. 

\subsection{TF-IDF} \label{sec:tf-idf}

So far we have compiled the most relevant terms in from the reviews. We now need to weight these terms for each review, so that we know the car-speak terms are most associated with a car. Using TF-IDF (Term Frequency-Inverse Document Frequency) has been used as a reliable metric for finding the relevant terms in a document~\cite{ramos2003}. 

We represent each review as a vector of TF-IDF scores for each word in the review. The length of this vector is $10,867$. We label each review vector with the car it reviews. We ignore the year of the car being reviewed and focus specifically on the model (i.e Acura ILX, not 2013 Acura ILX). This is because there a single model of car generally retains the same characteristics over time~\cite{yamawaki2002,lansley2016}.

\subsection{Classification Experiments}

We train a series of classifiers in order to classify car-speak. We train three classifiers on the review vectors that we prepared in Section~\ref{sec:tf-idf}. The classifiers we use are K Nearest Neighbors (KNN), Random Forest (RF), Support Vector Machine (SVM), and Multi-layer Perceptron (MLP)~\cite{scikitlearn2011}.

\begin{table}[h]
    \caption{Evaluation metrics for all classifiers.}
    \label{table:f1-scores}
    \begin{tabular}{| c | c | c | c | c |}
        \hline  
        & \textbf{KNN} & \textbf{RF} & \textbf{SVM} & \textbf{MLP} \\ 
        \hline
        \textbf{Precision Macro} & \textbf{0.6133} & 0.5968 & 0.6080 & 0.6094 \\ 
        \hline
        \textbf{Recall Macro} & \textbf{0.6086} & 0.5947 & 0.605 & 0.6059 \\ 
        \hline
        \textbf{F1 Macro} & \textbf{0.5808} & 0.5733 & 0.5801 & 0.5795 \\ 
        \hline
        \textbf{F1 Micro} & 0.6762 & 0.6687 & 0.6712 & \textbf{0.6778} \\ 
        \hline
    \end{tabular}
\end{table}

In order to evaluate our classifiers, we perform 4-fold cross validation on a shuffled data set. Table~\ref{table:f1-scores} shows the F1 micro and F1 macro scores for all the classifiers. The KNN classifier seem to perform the best across all four metrics. This is probably due to the multi-class nature of the data set. 


\section{Conclusion \& Future Work}

In this paper we aim to provide an introductory understanding of car-speak and a way to automate car dealers at dealerships. We first provide a definition of ``car-speak'' in Section~\ref{sec:car-speak}. We explore what constitutes car-speak and how to identify car-speak. 

We also gather a data set of car-speak to use for exploration and training purposes. This data set id full of vehicle reviews from U.S. News~\cite{usnews}. These reviews provide a reasonable set of car-speak data that we can study. 

Finally, we create and test several classifiers that are trained on the data we gathered. While these classifiers did not perform particularly well, they provide a good starting point for future work on this subject. 

In the future we plan to use more complex models to attempt to understand car-speak. We also would like to test our classifiers on user-provided natural language queries. This would be a more practical evaluation of our classification. It would also satisfy the need for a computer system that understands car-speak. 

\bibliographystyle{ACM-Reference-Format}
\bibliography{sample-bibliography}

\end{document}